\documentclass[9pt]{extarticle}
\usepackage[T1]{fontenc}
\usepackage{spconf,amsmath,graphicx}
\usepackage{cite}
\usepackage{amssymb,amsfonts}
\usepackage{graphicx,color}
\usepackage{textcomp}
\usepackage{xcolor}
\usepackage{hyperref}
\usepackage{relsize}

\usepackage{algorithm}
\usepackage{amsmath,bm,mathtools,mathrsfs,mathabx}
\usepackage{multicol}
\usepackage{amsthm}
\usepackage{float}
\usepackage{comment}

\usepackage{cleveref}
\usepackage{enumitem}
\usepackage{outlines}
\usepackage{algpseudocode}
\usepackage[skip=0pt]{caption}
\usepackage[skip=0pt]{subcaption}

\makeatletter

\theoremstyle{plain}
\newtheorem{thm}{\protect\theoremname}
\theoremstyle{plain}
\newtheorem{lem}[thm]{\protect\lemmaname}

\usepackage{bm}
\usepackage{babel}
\usepackage{amsmath}
\usepackage{bbm}
\usepackage{enumitem}
\usepackage{graphicx}
\usepackage{standalone}
\usepackage{booktabs}
\usepackage{multirow}
\usepackage{xcolor}
\usepackage{pdfcomment}
\usepackage{environ}
\usepackage{comment}

\usepackage{tikz}
\usetikzlibrary{arrows.meta,calc,positioning,shapes,fit}
\usepackage{pgfplots}
\pgfplotsset{compat=1.18}
\usepackage{standalone}

\hypersetup{colorlinks,breaklinks,
            linkcolor=[rgb]{0.5,0,0},
            citecolor=violet}

\newtheorem{assumption*}[assumption]{Assumption}

\newtheorem{insight*}[insight]{Insight}

\RenewEnviron{comment}{\pdfcomment{\BODY}}
\makeatother

\newcommand*{\hermtr}{{\mathsf{H}}}
\providecommand{\lemmaname}{Lemma}
\providecommand{\theoremname}{Theorem}

\renewenvironment{thebibliography}[1]{%
   \begin{oldthebibliography}{#1}%
     \setlength{\itemsep}{0ex}%
}%
{%
   \end{oldthebibliography}%
}

\title{Ultra-Reliable Risk-Aggregated Sum Rate Maximization via\\Model-Aided Deep Learning \vspace{-9pt}}
\name{Hassaan Hashmi$^\dagger$, Spyridon Pougkakiotis$^\ddagger$, and Dionysis Kalogerias$^\dagger$
\vspace{-10pt}}
\address{$^\dagger$Department of Electrical and Computer Engineering, Yale University, New Haven, USA\\
$^\ddagger$Department of Mathematics, King’s College London, London, England, UK\\
 \small{\tt \href{mailto:hassaan.hashmi@yale.edu}{hassaan.hashmi@yale.edu}, \href{mailto:spyridon.pougkakiotis@kcl.ac.uk}{spyridon.pougkakiotis@kcl.ac.uk}, \href{mailto:dionysis.kalogerias@yale.edu}{dionysis.kalogerias@yale.edu}}
\vspace{-14pt} \thanks{This work is supported by the US NSF under Grants 2242215 and 2431860.}.}

\begin{document}

\maketitle
\begin{abstract}
\vspace{-4pt}
We consider the problem of maximizing weighted sum rate in a multiple‐input single‐output (MISO) downlink wireless network with emphasis on user rate reliability. We introduce a novel risk-aggregated formulation of the complex WSR maximization problem, which utilizes the Conditional Value-at-Risk (CVaR) as a functional for enforcing rate (ultra)-reliability over channel fading uncertainty/risk. We establish a WMMSE-like equivalence between the proposed precoding problem and a weighted risk-averse MSE problem, enabling us to design a tailored unfolded graph neural network (GNN) policy function approximation (PFA), named \emph{$\alpha$-Robust Graph Neural Network ($\alpha$RGNN)}, trained to maximize \emph{lower-tail} (CVaR) rates resulting from adverse wireless channel realizations (e.g., deep fading, attenuation). We empirically demonstrate that a trained $\alpha$RGNN fully eliminates per user deep rate fades, and substantially and optimally reduces statistical user rate variability while retaining adequate ergodic performance.
\end{abstract}
\vspace{-5pt}
\begin{keywords}Beamforming, Weighted Sum Rate, Conditional Value-at-Risk (CVaR), Risk-Aware Optimization, Deep Unfolding.
\end{keywords}
\vspace{-10pt}
\section{Introduction} \label{sec: intro}
\vspace{-9pt}

\par Weighted sum rate (WSR) maximization is a well-studied and instructive problem in wireless systems engineering. Traditionally, the WSR problem is addressed either deterministically using well-known methodologies such as Zero-Forcing~\cite{zfbf}, WMMSE~\cite{wmmse}, or fractional programming~\cite{ccfp}, or stochastically through the design of ergodic-optimal resource allocation policies, maximizing expected WSR under stochastic resource constraints \cite{ergodic_stoch:kalogerias2020model}, under either perfect or imperfect channel state information (CSI)~\cite{lu2019robust:imperfectcsi}.

\par While resource policies (e.g., beamforming) optimizing ergodic (i.e., expected) or even deterministic QoS metrics (e.g., WSR) may be proven to 
perform optimally ``on average'' or ``in the long term'', they often result in \textit{inferior user-perceived system performance}. In particular, such optimal policies do not respond adequately to the presence of relatively (in)frequent, albeit operationally significant deep-fade events or, more generally, fading channel adversities, causing severe and abrupt drops in (perceived) service. This is a real and practical issue, especially considering that, in many actual scenarios, statistical dispersion of channel fading  typically exhibits heavy-tailed characteristics. 
In fact, it is well-known that ergodic-optimal policies often behave in a \textit{channel-opportunistic} manner~\cite{yoo2006optimality:opportunistic, ribeiro2012optimal:opportunistic}, completely discontinuing service to certain users in case of adverse channel conditions and low signal-to-noise ratio. This is not only inefficient in terms of communications, but also leads to substantial spectrum under-utilization.
  
\par To mitigate those issues, contemporary works have considered formulations based on outage probabilities, or explicit minimum user rate constraints 
\cite{dro:li2014,dro:li2017distributionally, bazzi2023:outageprob, alsenwi2023cvar:mincvarrate, li2024hpe:minqoscstr}. However, except for arguably being inappropriate for quantifying robustness and ensuring reliability, such objective/constraint choices are \textit{counterfactual by design}, since they rely on a-priori knowledge and/or choice of thresholds on (probabilistic) performance which are most often unavailable, and may be unachievable to begin with. For instance, how should we choose feasible
outage probability targets? And even if these targets are
feasible, how can we ensure they drive meaningful system improvements, like maintaining a sufficiently low
outage probability? Quantile-based measures, such as outage rates and capacities \cite{outagecap:lifang2001} in fact aim to address these issues. However, they again face serious
limitations in terms of effectiveness and interpretability, while lacking desirable technical properties such as convexity,
especially valuable in resource policy optimization.

More recently, a new line of work has been steadily emerging \cite{risk:bennis2018, risk:li2021, risk:vu2018, dro:yaylali2023, dro:yaylali2024}, where system robustness and/or reliability is quantified through the strategic use of \textit{risk measures} \cite{cvar:shapiro}, which are functionals genuinely generalizing expectations and designed to capture the statistical behavior (e.g., dispersion or volatility) of the random variables they operate with superior effectiveness, while also maintaining analytically attractive structure. Following this trend,
\textit{in this paper} we introduce the first \textit{risk-aggregated} formulation of the \textit{complex} WSR maximization problem, which utilizes the \textit{Conditional Value-at-Risk (CVaR)}~\cite{cvar:rockafellar} as a \textit{reliability functional} for quantifying performance (rate) over channel fading uncertainty/risk. CVaR is a \emph{coherent} risk measure~\cite{cvar:shapiro}, is free of counterfactual issues (see above), and interpolates seamlessly, interpretably and at will between the (maximally conservative) maximin (sometimes called ``robust'') and (naive) ergodic problem settings. These virtues makes CVaR an excellent performance measure for our purposes. Our contributions can be itemized as follows:
\vspace{-5pt}
\begin{itemize}[leftmargin=*]
    \item By enforcing optimal CVaR-based \textit{risk sharing among users} via risk aggregation, we enable the principled and systematic design of \textit{ultra-reliable, robust transmit precoding/beamforming policies}, beyond conventional ergodic/deterministic approaches.

    \vspace{-7pt}
    \item \sloppy We establish a WMMSE-like equivalence \cite{wmmse} between the risk-averse precoding problem and a weighted risk-averse MSE problem, exposing a splitting structure over the decision variables. This informs the design of a tailored, \textit{model-aided deep unfolded GNN-based PFA} 
    called \textit{$\alpha$RGNN}, resembling the iterative structure of block-coordinate methods, akin to WMMSE. Instead of employing monolithic black-box Deep Neural Networks (DNNs), we are able to curb over-parameterization and reduce complexity.
    \vspace{-7pt}
    \item We demonstrate that a well-trained $\alpha$RGNN is able to eliminate opportunistic policy behavior due to channel fading adversities (e.g., deep fades or attenuation), while optimally reducing the statistical variability of user-experienced rates. In fact, $\alpha$RGNN achieves optimal and fully tunable system (ultra-)reliability,
    drastically improving over established techniques, such as WMMSE.
\end{itemize}
\vspace{-5pt}
\textit{Note:} Proofs are omitted, to be included in an extended journal submission. Code is available at: \href{https://github.com/hassaanhashmi/argnn}{github.com/hassaanhashmi/argnn}.

\vspace{-4pt}
\section{Problem Formulation}
\vspace{-7pt}
In this work, we consider a MISO wireless network cell with a base station (BS) having $M$ antennas serving a set of $\mathcal{I}=\{1,2, \ldots, K\}$ users. Each user's bit rate at the downlink is defined as
\begin{equation}
r_{i}\left(\bm{V},\bm{h}_{i}\right) \triangleq \log\left(1+ \frac{\left|\bm{h}_i^\hermtr\bm{v}_i\right|^2}{{\sum}_{{j \in \mathcal{I} \setminus i}}\left|\bm{h}^\hermtr_i\bm{v}_j\right|^2 + \sigma^2_i}\right),\quad\forall\ i\in\mathcal{I},\label{eq:rate}
\end{equation}
where $\sigma_i^2$ is the noise variance, $\bm{h}_{i} {\in} \mathbb{C}^{M\times1}$ is the random fading channel of the corresponding link between the $i$-th user and the BS, and $\bm{V} {=} [\bm{v}_1 \,\bm{v}_2 \,{\cdots} \,\bm{v}_K ]\ {\in}\ \mathbb{C}^{M \times K}$ are the precoding vectors.
Given rates in \eqref{eq:rate} and a total allocated power of $P_{BS}$, the \textit{expected} weighted sum rate (WSR) maximization problem is formulated as
\begin{equation} \label{eq:expected_wsr_pf}
\max_{\|\bm{V}(\cdot)\|^2_F \leq P_{BS}}\sum_{i\in\mathcal{I}}\gamma_{i}\mathbb{E}\left\{ r_{i}\left(\bm{V}\left(\bm{H}\right),\bm{h}_{i}\right)\right\} 
\end{equation}
where $\bm{H} {=} \left[\begin{array}{@{}c@{}c@{}c@{}c@{}} \bm{h}_1 \: & \bm{h}_2 \: & {\cdots} \: & \bm{h}_K \end{array}\right] {\in}\ \mathcal{H} \subseteq \mathbb{C}^{M \times K}$ is the corresponding channel matrix, $\bm{V}\colon\mathcal{H}\to\mathbb{C}^{M \times K}$ is the precoding policy, and each $\gamma_{i}$ is a
predetermined constant. The constraint $\|\bm{V}(\cdot)\|_F^2\leq P_{BS}$ (where $\|\cdot\|_F$ denotes the Frobenius norm) forces the precoders to satisfy the power constraint, with maximum allowed power $P_{BS}$.

\par As mentioned in Section \ref{sec: intro}, an inherent issue with this formulation is that it is unable to capture the statistical volatility of the user rates. Thus, the resulting ergodic-optimal policies may behave opportunistically (in fact, they do) to increase expected WSR performance by shutting off users with deep faded, or more generally weak, channels, or low singal-to-noise ratios.

\par To enable sensitivity to high-order statistical behavior of the rate distributions in the WSR problem, we introduce an alternative model utilizing the \textit{$\alpha$-percentile Conditional Value-at-Risk ($\text{CVaR}_{\alpha}$)}~\cite{cvar:rockafellar}, which, for an integrable random \textit{cost} $Z$, is  defined as
\begin{equation}
\text{CVaR}_{\alpha}\left(Z\right) \triangleq \inf_{t\in\mathbb{R}}\:t + \alpha^{-1}\mathbb{E}\left\{ \left(Z-t\right)_{+}\right\} \label{eq:cvar}
\end{equation}
where $\alpha=(0, 1]$ and $(\cdot)_+ \triangleq \max\{\cdot,0\}$. The $\text{CVaR}_{\alpha}$ expresses \textit{the mean of the worst}
$\alpha \cdot 100 \%$ values of $Z$, while generalizing the expectation in a strict and tractable sense, as
\begin{align}
\begin{aligned}\label{eq:cvar_exp}
\text{CVaR}_{1}\left(Z\right) &= \mathbb{E}\{Z\} \leq \text{CVaR}_{\alpha}\left(Z\right), \ \forall\ \alpha \in (0, 1], \:\: \text{and}
\\
\text{CVaR}_{0}\left(Z\right) &\triangleq \lim_{\alpha\downarrow 0} \text{CVaR}_{\alpha}\left(Z\right) =  \text{ess}\sup Z. 
\end{aligned}
\end{align}

\par
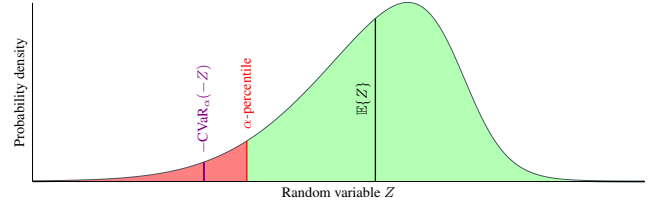
\begin{figure}[!t]
  \centering

\pgfmathdeclarefunction{gauss}{2}{%
  \pgfmathparse{exp(-( (#1)^2 )/(2*#2^2))/(#2*sqrt(2*pi))}%
}

\pgfmathdeclarefunction{skewnormal}{3}{%
  \pgfmathparse{2*gauss(#1,#3) * (1/(1+exp(-#2*(#1/#3))))}%
}

\begin{tikzpicture}
  \begin{axis}[
      no markers,
      domain=-7:3,
      samples=200,
      axis lines*=left,
      height=5.5cm,
      width=15cm,
      xtick=\empty,
      ytick=\empty,
      xlabel={Random variable $Z$},
      ylabel={Probability density},
      enlargelimits=false,
      clip=false,
      axis on top,
    ]
    \addplot [very thick, cyan!20!black] {skewnormal(x,-5,1.9)};

    \addplot [
      draw=none,
      fill=red!50,
      domain=-7:-3.5
    ] {skewnormal(x,-5,1.9)} \closedcycle;

    \addplot [
      draw=none,
      fill=green!30,
      domain=-3.5:3
    ] {skewnormal(x,-5,1.9)} \closedcycle;

    \addplot[black, thick] coordinates {(-1.4,0) (-1.4,0.313)};
    \node[black, rotate=90, above] at (axis cs:-1.4, 0.16) {$\mathbb{E}\{Z\}$};

    \addplot[red,  thick] coordinates {(-3.5,0) (-3.5,0.078)};
    \node[red, rotate=90, above] at (axis cs:-3.3, 0.17) {$\alpha$-percentile};

    \addplot[violet, very thick] coordinates {(-4.2,0) (-4.2,0.037)};
    \node[violet, rotate=90, above] at (axis cs:-4.0, 0.14) {$-\text{CVaR}_\alpha(-Z)$};

  \end{axis}
\end{tikzpicture}
  \caption{A depiction of $-\text{CVaR}_{\alpha}(-Z)$}
  \label{fig:minus_cvar_minus}
  \vspace*{-15pt}
\end{figure}

 Problem \eqref{eq:expected_wsr_pf} maximizes rewards rather than minimizing costs. Therefore, we accordingly \textit{reflect} $\text{CVaR}_{\alpha}$ as
\begin{equation}
-\text{CVaR}_{\alpha}\left(-Z\right)=\sup_{t\in\mathbb{R}}\:t-\alpha^{-1}\mathbb{E}\left\{ \left(t-Z\right)_{+}\right\}, \label{eq:cvar_minus}
\end{equation}
now measuring expected value restricted to the \emph{lower} tail of the \textit{reward} $Z$, as shown in Figure \ref{fig:minus_cvar_minus}.
\par The proposed risk-aggregated (sum-of- $\text{CVaR}_{\alpha}$s) WSR problem reads
\begin{equation} \label{eq:cvar_wsr}
 \boxed{\max_{\|\bm{V}(\cdot)\|_F^2 \leq P_{BS}}-\sum_{i\in\mathcal{I}}\gamma_{i}\text{CVaR}_{\alpha_{i}}\left\{ -r_{i}\left(\bm{V}\left(\bm{H}\right),\bm{h}_{i}\right)\right\},}
\end{equation}
where each $\alpha_i$ denotes the user-specific percentile level. We emphasize that, unlike problem \eqref{eq:expected_wsr_pf}, the risk-aware formulation in \eqref{eq:cvar_wsr} does \textit{not} admit an equivalent deterministic formulation (in light of interchangeability principle~\cite[Section 9.3.4]{cvar:shapiro}), exactly due to the presence  of risk aggregation.

\vspace{-4pt}
\section{An Equivalent WMMSE-like Formulation}
\vspace{-7pt}
\par The WMMSE algorithm~\cite{wmmse} is a classical block-coordinate scheme used to approximately solve the non-convex WSR maximization problem in a pointwise fashion by  utilizing an equivalent problem reformulation with auxiliary variables, which are used to derive three associated tractable subproblems (at each WMMSE iteration). In that framework, the mean squared error (MSE) between a sent and received symbol for the $i$-th user is given by~\cite[Eqn. (3)]{wmmse} \begin{align} \label{eq: mse}
\hspace{-4pt}e_{i}(u_{i},&\bm{V},\bm{h}_{i})\hspace{-1pt}=\hspace{-2pt} 
 \left|1 \hspace{-1pt}-\hspace{-1pt}u_{i}^{*}\bm{h}_{i}^{\hermtr}\bm{v}_{i}\right|^{2}
 \hspace{-1pt}+\hspace{-1pt}
 \sum_{j\in\mathcal{\mathcal{I}}\setminus i}\left|u_{i}^{*}\bm{h}_{i}^{\hermtr}\bm{v}_{j}\right|^{2}+\sigma_{i}^{2}\left|u_{i}\right|^{2},\hspace{-4pt}
\end{align}
where $u_i$ is a scalar receive beamformer under the MISO setup and $(\cdot)^*$ denotes the complex conjugate.

\par Drawing inspiration from the development of WMMSE (which tackles the WSR problem \eqref{eq:expected_wsr_pf}), one can investigate the possibility of designing a corresponding block-coordinate scheme for problem \eqref{eq:cvar_wsr} (akin to WMMSE). To that end, we first expand \eqref{eq:cvar_wsr} as
\begin{equation} \nonumber
    \max_{\|\bm{V}(\cdot)\|_F^2 \leq P_{BS}}  \max_{\bm{t} \in \mathbb{R}^{K}} \mathbb{E}\left\{ \sum_{i\in\mathcal{I}}\gamma_{i}\left[t_{i}-\alpha_{i}^{-1}\left(t_{i}-r_{i}\left(\bm{V}\left(\bm{H}\right),\bm{h}_{i}\right)\right)_{+}\right]\right\}.
\end{equation}
We may establish the following equivalence, a la  WMMSE.

\begin{lem}
\label{lem:wmmse_reformulation}
For every $\bm{V}\left(\cdot\right)$, $\bm{H} \in \mathcal{H}$, and $t_{i}\in\mathbb{R}$, $i\in\mathcal{I}$, it holds that
\begin{align*}
&t_{i}-\alpha_{i}^{-1}\left(t_{i}-r_{i}\left(\bm{V}(\bm{H}),\bm{h}_{i}\right)\right)_{+}=\nonumber \\ 
&\max_{u_{i}\in\mathbb{C},w_{i} \in \mathbb{R}} \left[t_{i}-\alpha_{i}^{-1}\left(t_{i} -R_{i}\left(u_{i},w_{i},\bm{V}(\bm{H}),\bm{h}_{i}\right)\right) \right], \nonumber
\end{align*}
\noindent where $R_{i}\left(u_{i},w_{i},\bm{V}(\bm{H}),\bm{h}_{i}\right)\triangleq\log w_{i}-w_{i}e_{i}\left(u_{i},\bm{V}(\bm{H}),\bm{h}_{i}\right)$.
\end{lem}

%
By using the interchangeability principle~\cite[Section 9.3.4]{cvar:shapiro}, we can now rewrite the nonconvex problem~\eqref{eq:cvar_wsr}
as
\begin{equation}\label{eq:wmmse_equiv}
\max_{\bm{t}\in\mathbb{R}^{K}}\mathbb{E}
\hspace{-1pt}
\left\{ \max_{\substack{ \bm{u},\bm{w}  \\ \|\bm{V}\|_F^2 \leq P_{BS}} }\sum_{i\in\mathcal{I}}\gamma_{i}\left[t_{i}-\alpha_{i}^{-1}\left(t_{i}{-}R_{i}\left(u_{i},w_{i},\bm{V},\bm{h}_{i}\right)\right)_{+}\right]\right\}, 
\end{equation}
where $\bm{u} {=} [ u_1 \, u_2 \, {\cdots} \, u_K ]^\intercal$ and $\bm{w} {=} [ w_1 \,  w_2 \,  {\cdots} \,  w_K]^\intercal$. The problem inside the expectation in \eqref{eq:wmmse_equiv} resembles the classical formulation of WMMSE, but does not admit an efficient block-coordinate scheme akin to WMMSE, because of the presence of $(\cdot)_+$.

\begin{algorithm}[!t]
\caption{Training of $\alpha$RGNN with SsGA}
\label{Algorithm: aRGNN}
\begin{algorithmic}
\State \textbf{Input: }$\bm{\theta}_{0}=\left[\bm{\theta}_{u,0}, \bm{\theta}_{w,0},\bm{\theta}_{v,0}\right]$, $\bm{t}_{0}=\left\{ t_{i,0}\right\} _{i\in\mathcal{I}}$,$\tau=0$; $\left\{ \gamma_{i}\right\} _{i\in\mathcal{I}}$,
$\left\{ \alpha_{i}\right\} _{i\in\mathcal{I}}$,$0<\eta_{\bm{\theta}}^0,\eta_{t}^0\ll1$ , $T>0$,
$L>1$

\For {$\tau \leq T$}

\State Sample i.i.d. $\bm{H}\in\mathcal{H}$

\State Initialize $\bm{V}^{\left(0\right)}_{\bm{z}_{\bm{\theta}}}$ such that $\sum_{i\in\mathcal{I}}\|\bm{v}^{\left(0\right)}_{{\bm{z}_i}}\|^{2}\leq P_{\text{BS}}$

\For {$\ell=1,\ldots,L$}

\State Update each $\hat{\bm{u}}_{{i}}^{\left(\ell\right)}$ by \eqref{eq:ugnn}, $\forall\ i\in\mathcal{I}$

\State Update each $\hat{\bm{w}}_{{i}}^{\left(\ell\right)}$ by \eqref{eq:wgnn}, $\forall\ i\in\mathcal{I}$

\State Update each $\hat{\bm{v}}_{{i}}^{\left(\ell\right)}$ by \eqref{eq:vgnn}, $\forall\ i\in\mathcal{I}$

\State Project each $\hat{\bm{v}}_{{i}}^{\left(\ell\right)}$ to $\tilde{\bm{v}}_{{i}}^{\left(\ell\right)}$
by (\ref{eq:cstr_projection}), $\forall\ i\in\mathcal{I}$
\EndFor

\State Choose $\eta_{\bm{\theta}}^{\tau}$ and $\eta_{t}^{\tau}$ and, $\forall i \in \mathcal{I}$, set
\vspace{-5pt}
$$\bm{\theta}_{\tau+1}{\leftarrow}\bm{\theta}_{\tau}{-}\eta_{\theta}^{\tau}\nabla_{\bm{\theta}}\sum_{i\in\mathcal{I}}\frac{\gamma_{i}}{\alpha_i}\left(t_{i,\tau}-r_{i}\left(\tilde{\bm{V}}_{\bm{z}_{\bm{\theta}}}^{\left(L\right)},\bm{h}_{i}\right)\right)_{+}$$ \\
\vspace{-13pt}
$$t_{i,\tau+1} \leftarrow t_{i,\tau} {+} \eta_{t}^{\tau}\frac{\gamma_{i}}{\alpha_i} \left(\alpha_i{-}\nabla_{t_i}\left(t_{i}-r_{i}\left(\tilde{\bm{V}}_{\bm{z}_{\bm{\theta}}}^{\left(L\right)},\bm{h}_{i}\right)\right)_{+}\right)$$
\State where the (sub)gradients are computed via autodifferentiation
\EndFor
\end{algorithmic}
\end{algorithm}

\par Instead, we establish equivalence between \eqref{eq:cvar_wsr} and \eqref{eq:wmmse_equiv} to inform the development of a WMMSE-like \textit{recurrent} policy function approximation (PFA), resembling the structure of a (deep-unfolded) block-coordinate scheme, \emph{to tackle the original problem \eqref{eq:cvar_wsr}}. In particular, and as will become clear later on, to avoid utilizing large unstructured PFAs, we take advantage of the structure of problem \eqref{eq:wmmse_equiv} and restrict the parametric space of the beamforming policies accordingly.

\vspace{-4pt}
\section{A Model-aided Graph Neural Network} \label{sec:gnn}
\vspace{-7pt}
As mentioned above, we now propose a \textit{model-aided~\cite{shlezinger2023model:MBDL} PFA}, by utilizing a graph neural network (GNN)~\cite{GNN1} architecture. In what follows, we first provide a graph data representation of a MISO downlink network. Subsequently, inspired by recent findings in~\cite{zhao2024understanding:cnngnncomp}, we present our WMMSE-like deep unfolding GNN-based PFA architecture, termed \textit{$\alpha$-Robust Graph Neural Network ($\alpha$RGNN)}.

\vspace{-6pt}
\subsection{Network Realizations as Graph Data} \label{subsec:graphdata}
\vspace{-5pt}
We first consider each MISO downlink network (channel) realization as a directed homogenous graph $\mathcal{G}\left(\cdot\right)=\left(\mathcal{V}\left(\cdot\right),\mathcal{E}\left(\cdot\right)\right)$, in which $\mathcal{V}\left(\cdot\right)=\left\{ \bm{z}_{i}\left(\cdot\right)\right\} _{i\in\mathcal{I}}$ are the set of vertices, where each node feature $\bm{z}_{i}\left(\cdot\right)$, for a particular realization of $\bm{H}$, is given as
\[
\bm{z}_{i}\left(\bm{h}_{i}\right)=\left[
(\hat{\bm{v}}^{(0)}_{{i}})^\intercal \,\, \bm{h}_{i}^{\intercal} \,\,
(\hat{\bm{u}}^{(0)}_{{i}})^\intercal \,\, (\hat{\bm{w}}^{(0)}_{{i}})^\intercal \,\,
\gamma_{i} \right]^{\intercal},\quad \forall\ i\in\mathcal{I},
\]
where $\hat{\bm{v}}^{(0)}_{{i}}{\in}\mathbb{C}^{M}$, $\hat{\bm{u}}^{(0)}_{{i}}{\in}\mathbb{C}^{d_{u}}$, and $\hat{\bm{w}}^{(0)}_{{i}}{\in}\mathbb{C}^{d_{w}}$ are the initial values of these vectors corresponding to each user $i$. To enhance the feature extraction capacity of the GNN (in latent space), we consider the vectors $\hat{\bm{u}} \in \mathbb{C}^{d_u}$ and $\hat{\bm{w}} \in \mathbb{R}^{d_w}_+$ in place of the scalars $u_i$ and $w_i$, respectively~\cite{yang2024knowledge:uwgnn}. Finally, the interference link from user $j$ to $i$ is represented as an edge between node $j$ and node $i$. Specifically, each edge in the adjacency tensor $\bm{A}\left(\cdot\right) \in \mathbb{C}^{K \times K \times M}$ is defined as
\[
\bm{A}_{\left(j,i, \cdot\right)}\left(\bm{H}\right)=\begin{cases}
\bm{0}^\intercal & \text{if }j\notin\mathcal{N}\left(i\right)\\
\bm{h}_{j}^\intercal & \text{otherwise}
\end{cases},
\]
where $\mathcal{N}\left(i\right)$ denotes the neighborhood of node $i$.

\vspace{-6pt}
\subsection{$\alpha$RGNN architecture}
\vspace{-5pt}
We propose a deep GNN-based model comprised of three cascading GNNs: uGNN, wGNN, and vGNN. First, the uGNN updates each node subfeature $\hat{\bm{u}}_{{i}}\in\mathbb{C}^{d_{u}}, \forall i\in\mathcal{I}$. Concretely, for the $\ell$-th unfolding layer, we have
\vspace{-3pt}
\begin{equation}\label{eq:ugnn}
\hat{\bm{u}}_i^{(\ell)} = \phi_u\Big( \bm{h}_i, \tilde{\bm{v}}_i^{(\ell-1)}{,} \bigoplus_{j \in \mathcal{N}(i)} \psi_u( \tilde{\bm{v}}_j^{(\ell-1)}; \bm{\theta}_{\psi_u} ); \bm{\theta}_{\phi_u} \Big),
\vspace{-6pt}
\end{equation}
\sloppy where $\psi_u\left(\cdot;\bm{\theta}_{\psi_u}\right)$ outputs message embeddings from the neighborhood of a node, $\bigoplus_{j \in \mathcal{N}(i)}$ represents message aggregation over those embeddings, and $\phi_u\left(\cdot;\bm{\theta}_{\phi_u}\right)$ updates the values of node subfeature $\hat{\bm{u}}_{{i}}$ respectively, for every node (user). Here, each $\tilde{\bm{v}}_i^{(\ell-1)}$ is a projection of $\hat{\bm{v}}_i^{(\ell-1)}$ onto a feasible set subject to the power constraint. This implies that each $\hat{\bm v}^{(0)}_{i}$ must be feasible from the outset and, as discussed in the next section, that is indeed the case.

\par Next, for the same $\ell$-th unfolding, we update each $\hat{\bm{w}}_{{i}}\in\mathbb{C}^{d_{w}}$ with the wGNN, the architecture of which is given by
\vspace{-2pt}
\begin{equation}\label{eq:wgnn}
\hat{\bm{w}}_i^{(\ell)} {=} \phi_w\Big( \bm{h}_i,  \tilde{\bm{v}}_i^{(\ell-1)}, \hat{\bm{u}}_i^{(\ell)}{,} \bigoplus_{j \in \mathcal{N}(i)} \psi_w( \tilde{\bm{v}}_j^{(\ell-1)}; \bm{\theta}_{\psi_w} ); \bm{\theta}_{\phi_w} \Big),
\vspace{-6pt}
\end{equation}
where, for each node $i$, the $\hat{\bm{u}}_i^{(\ell)}$ updated by uGNN is utilized by wGNN for the same node $i$.
\begin{figure}[t]
  \centering
  \centerline{\includegraphics[width=3in]{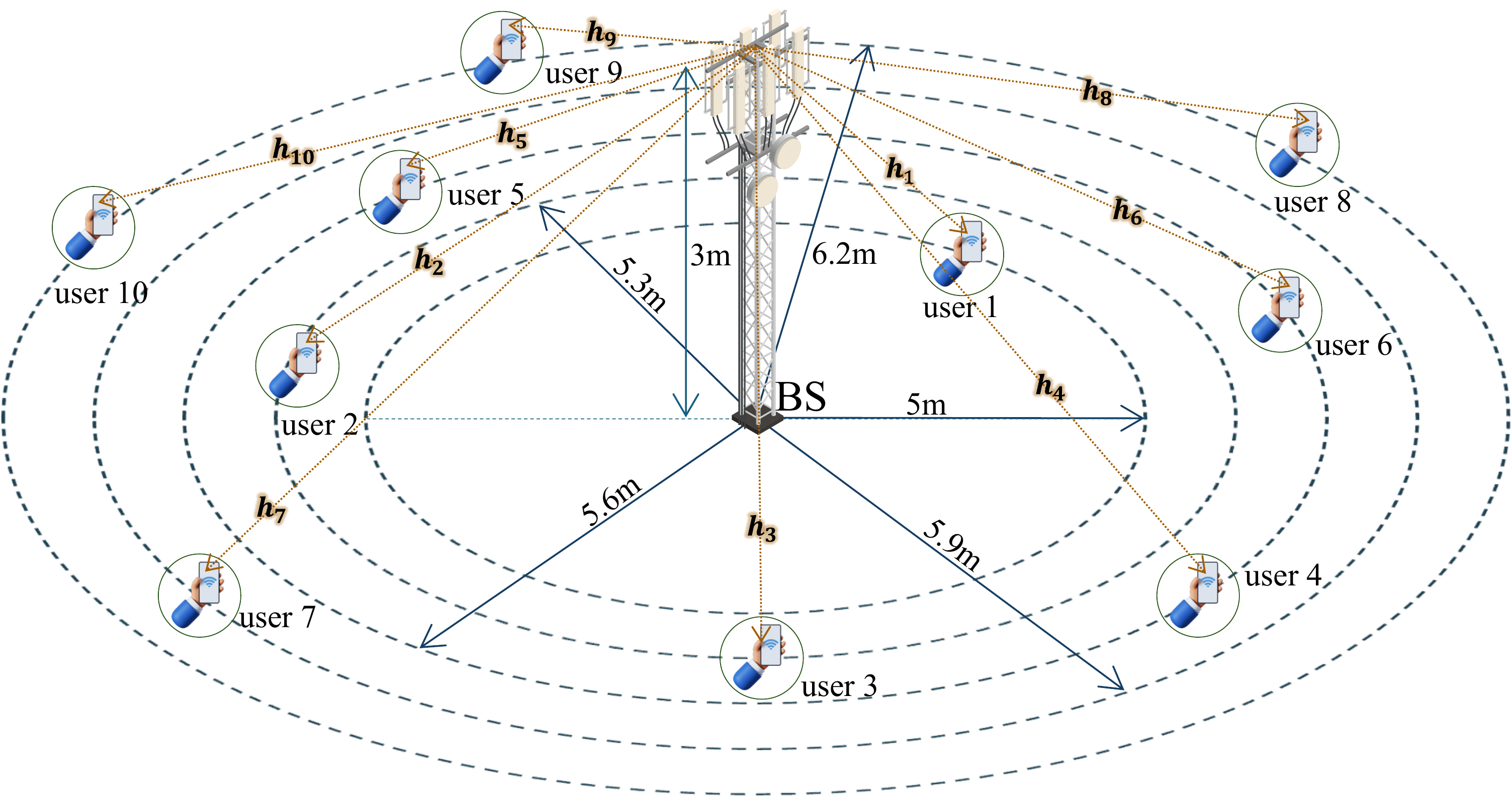}}
\vspace{6pt}
\caption{Simulated MISO downlink network configuration.}
\label{fig:env_setup}
\vspace{-12pt}
\end{figure}
\par Then, given the updated $\{ \hat{\bm{u}}_{{i}}^{\left(\ell\right)},\hat{\bm{w}}_{{i}}^{\left(\ell\right)}\} _{i\in\mathcal{I}}$, we further update every $\hat{\bm{v}}_{{i}}^{\text{\ensuremath{\left(\ell-1\right)}}}$ by the vGNN architecture as
\begin{equation}
\begin{split}\label{eq:vgnn}
\bm{\xi}^{\left(\ell\right)}_{j} &= \psi_v\left(
\bm{h}_{j},\tilde{\bm{v}}_{j}^{\left(\ell-1\right)}, 
\hat{\bm{u}}_{j}^{\left(\ell\right)},\hat{\bm{w}}_{j}^{\left(\ell\right)};\bm{\theta}_{\psi_v}
\right), \\ 
\hat{\bm{v}}_{{i}}^{\left(\ell\right)} &= \phi_{v}\Big(\bm{h}_{i},\tilde{\bm{v}}_{{i}}^{\left(\ell-1\right)},\hat{\bm{u}}_{{i}}^{\left(\ell\right)},\hat{\bm{w}}_{{i}}^{\left(\ell\right)},\bigoplus_{j \in \mathcal{N}(i)}\bm{\xi}^{\left(\ell\right)}_{j};\bm{\theta}_{\phi_v}\Big),
\end{split}
\end{equation}
where $\hat{\bm{v}}_{{i}}^{\left(\ell\right)}$ may no longer be feasible. 
We may then \textit{trainably (and implicitly) constrain} every $\hat{\bm{v}}_i$ by setting
\begin{equation} \label{eq:cstr_projection}
\tilde{\bm{v}}_{i}^{\left(\ell\right)}=\begin{cases}
\hat{\bm{v}}_{{i}}^{\left(\ell\right)} & \text{if }\sum_{i\in\mathcal{I}}\|\hat{\bm{v}}_{{i}}^{\ell}\|^{2}\leq P_{\text{BS}}\\
\hat{\bm{v}}_{{i}}^{\left(\ell\right)}\times\sqrt{\frac{P_{\text{BS}}}{\sum_{i\in\mathcal{I}}\|\hat{\bm{v}}_{{i}}^{\ell}\|^{2}}} & \text{if }\sum_{i\in\mathcal{I}}\|\hat{\bm{v}}_{{i}}^{\ell}\|^{2}>P_{\text{BS}}
\end{cases}.
\end{equation}
\begin{figure*}[t]
  \centering
  \includegraphics[width=\textwidth]{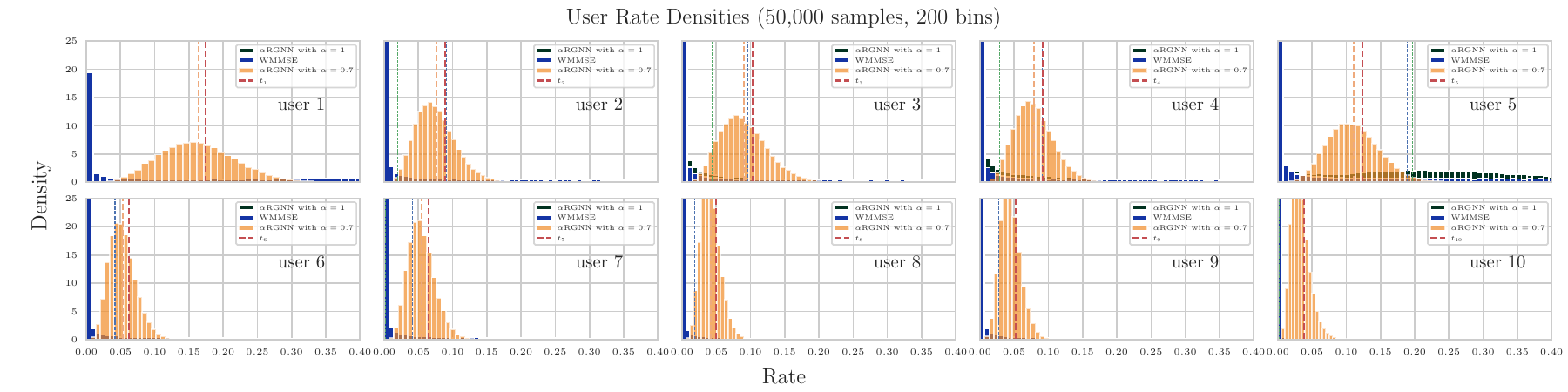}
\vspace{-8bp}
\caption{User rate densities achieved by WMMSE~\cite{wmmse}, \emph{risk-neutral} $\alpha$RGNN, and $\alpha$RGNN with $\alpha=0.7$ on 50,000 channel realizations and 200 bins; vertical dotted lines represent sample averages (rate and density axes ranges are kept consistent for all users).
}
\label{fig:histogram}
\vspace{-12bp}
\end{figure*}

The GNNs (\ref{eq:ugnn}), (\ref{eq:wgnn}), (\ref{eq:vgnn}), and the projection \eqref{eq:cstr_projection}, constitute one complete unfolding layer of the $\alpha$RGNN, which we may use to 
formulate problem 
\begin{equation} \label{eqn:cvar_surrogate}
    \max_{\bm{\theta}, \bm{t}} \mathbb{E}\left\{ \sum_{i\in\mathcal{I}}\gamma_{i}\left[t_{i}-\alpha_{i}^{-1}\left(t_{i}-r_{i}\left(\tilde{\bm{V}}^{(L)}_{\bm{z}_{\bm{\theta}}}\left(\bm{H}\right),\bm{h}_{i}\right)\right)_{+}\right]\right\},
\end{equation}

\noindent on which we apply a stochastic subgradient scheme to optimize over $\bm{\theta}$ and $\bm{t}$ (i.e., train), as shown in Algorithm \ref{Algorithm: aRGNN},  where the evaluation of $\tilde{\bm{V}}^{(L)}_{\bm{z}_{\bm{\theta}}}{=} [\tilde{\bm{v}}^{(L)}_1 \, \tilde{\bm{v}}^{(L)}_2 \, {\cdots} \, \tilde{\bm{v}}^{(L)}_K ]$ on $\bm{H}$ can be viewed as the \emph{action} of the $\alpha$RGNN \emph{policy}, and $L$ is the corresponding unfolding depth.
\par Before closing this section, let us note that Algorithm \ref{Algorithm: aRGNN} can be shown to converge to an approximately stationary (in the \emph{Clarke} sense) point of \eqref{eqn:cvar_surrogate} under minimal assumptions, by applying the theory developed in~\cite{DavisDrusvyatskiyKakadeLee2020_TameSubgradient}. Notably, the core (but minor) assumption required by the aforementioned theory is that the objective function of \eqref{eqn:cvar_surrogate} is \emph{Whitney stratifiable}. A detailed discussion on the convergence properties of Algorithm \ref{Algorithm: aRGNN} are omitted here for brevity of exposition and the reader is referred to~\cite{DavisDrusvyatskiyKakadeLee2020_TameSubgradient,BolteLeMoulinesPauwels2024_InexactSubgradient} for additional details.

\vspace{-4pt}
\section{Simulations}
\vspace{-7pt}
\par In our simulations, we consider a MISO-downlink cell with $K{=}10$ users and a BS with $M{=}6$ antennas as shown in Figure \ref{fig:env_setup}. We assume Rician fading on all the channel links (with a Rician factor of $\beta{=}$-3dB), where the entries of the instantaneous CSI (I-CSI) are drawn i.i.d. from $\mathcal{CN}(0,1)$, while the entries of the statistical CSI (S-CSI) are drawn once from $\mathcal{CN}(0,1)$ and remain fixed throughout the operation of the BS. We assume $P_{BS}{=}$5dBm, while the channel noise variance for each user is set to $\sigma^2_i{=}$ -80dBm, $\forall\ i \in \mathcal{I}$. The pathloss of each link is modeled as $L=\sqrt{C_0 d^{-\delta}}$, where $d$ is the distance (in meters), $C_0{=}$-30dBm, and $\delta{=}$2.2. 

\par As in Section \ref{subsec:graphdata}, we construct a graph for each channel realization, where we consider both $d_u$ and $d_w$ to be 16, and initialize each $\hat{\bm{u}}_i$ and $\hat{\bm{w}}_i$ with all entries equal to zero. The beamformers $\bm{v}_i$ are initialized according to a uniform policy, subject to the power constraint, of course.

\par For the $\alpha$RGNN, the parameter vectors $\bm{\theta}_{\phi_u}$, $\bm{\theta}_{\psi_u}$, $\bm{\theta}_{\phi_w}$, $\bm{\theta}_{\psi_w}$, $\bm{\theta}_{\phi_v}$, and $\bm{\theta}_{\psi_v}$ are substituted by ReLU-activated 3-layer fully-connected neural networks (FNNs) with 256 neurons in each hidden layer, while the input and output dimensions follow directly from the domain and codomain dimensions specified in \eqref{eq:ugnn}, \eqref{eq:wgnn}, and \eqref{eq:vgnn}. For each message aggergation $\bigoplus_{j \in \mathcal{N}(i)}$, we take $\max_{j \in \mathcal{N}(i)}$. Finally, we take $L=4$, i.e., we unfold the model 4 times. It should also be mentioned here that for each complex-valued vector, say $\bm{y}\in \mathbb{C}^p$, we use the mapping $\bm{y} \mapsto \begin{bmatrix}\Re(\bm{y})^\intercal \Im(\bm{y})^\intercal \end{bmatrix}^\intercal \in \mathbb{R}^{2p}$ in both the graph data construction as well as in $\alpha$RGNN, where the input and output dimensions of the FNNs follow accordingly.

\par We first sample 200,000 i.i.d. channel realizations to create a graph dataset ($\gamma{=}1$ for each user), scaling each channel realization by its maximum magnitude entry for consistency during training. We then train two $\alpha$RGNN's, one with $\alpha{=}0.7$ for each user (risk-aware) and the other with each $\alpha{=}1$ (\emph{risk-neutral}), on the graph data for 40 epochs with a batch size of 64.  We choose initial learning rates $\eta^0_\theta{=}10^{-3}$ and $\eta^0_t{=}10^{-5}$ with a step decay of 0.794 and 0.912 per epoch after the first 5 epochs, respectively. 

\par We then evaluate the trained models, as well as the WMMSE method~\cite{wmmse} (using 20 internal iterations), on a test set of 50,000 graphs and their corresponding channel realizations, respectively, where WMMSE achieves a higher average sumrate (1.11 bits/s/Hz) than the risk-aware $\alpha$RGNN (0.74 bit/s/Hz). However, as shown in Figure \ref{fig:histogram}, ``zero rate'' events are essentially eliminated for all users when evaluating the risk-aware $\alpha$RGNN policy, in contrast to the \emph{risk-neutral} WMMSE and $\alpha$RGNN policies, which indeed show ergodic opportunistic behaviors and almost fully disrupt service to the distant users 8, 9, and 10 (see Figure \ref{fig:env_setup}) in favor of the nearby users. Further, each user's rate variance is reduced substantially under the risk-aware $\alpha$RGNN model. These findings demonstrate that our risk-aware formulation yields beamforming policies that are both robust and reliable (i.e., mitigating the occurrence of channel adversities while exhibiting reduced rate variance).

\begin{figure}
  \centering
  \centerline{\includegraphics[width=3.5in]{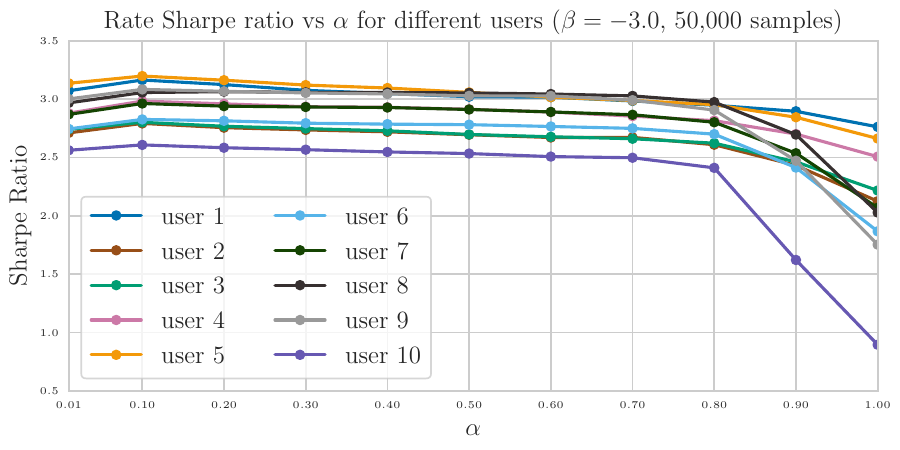}}
\vspace{1pt}
\caption{User Rate Sharpe ratios~\cite{sharperatio} achieved by $\alpha$RGNN.}
\label{fig:sharpe}
\vspace{-16pt}
\end{figure}

To further comment on the selection of $\alpha$‐percentiles, we train eleven $\alpha$RGNN models at different risk ($\alpha$-percentile) levels (see Figure \ref{fig:sharpe}). We then evaluate each model by computing, for each user, the Sharpe ratio~\cite{sharperatio} of the achieved rate at each $\alpha$‐percentile risk. The Sharpe ratio here is defined as the sample mean of a user’s rate divided by its standard deviation, serving as a metric of volatility and, by extension, reliability. The trend in Figure \ref{fig:sharpe} confirms that the risk‐neutral policy (i.e.\ $\alpha {=} 1$) exhibits the highest risk and correspondingly lowest reliability. We also note that the most conservative policy (at $\alpha {=} 0.01$) shows a decline in the Sharpe ratios, which is in line with expected behavior.

\vspace{-9pt}
\section{Conclusions} \label{sec: Conclusions}
\vspace{-9pt}
In this work, we proposed a novel risk-aggregated formulation for the WSR problem in MISO downlink wireless networks, utilizing the CVaR risk measure. Through an interesting WMMSE-like equivalence, we developed a model-based deep unfolding GNN, named $\alpha$RGNN, enabling the design of optimal ultra-reliable transmit precoding/beamforming policies, beyond conventional ergodic/deterministic approaches. Our empirical analysis corrobora- ted the efficacy of $\alpha$RGNN in achieving both robust and reliable rate performance on a per user basis. Even though this work was developed around a convenient MISO downlink setup, our approach can be extended to a wide range of settings to optimize user bit rates in a risk-aware fashion (akin to the general WMMSE setting \cite{wmmse}). Investigating risk-aggregation in other well-known QoS metrics, as well as a systematic exploration of different model-based PFA parameterizations, remain interesting avenues for future research.

\clearpage
\bibliographystyle{IEEEtran}
\bibliography{argnn_bib}

\begin{thebibliography}{10}
\providecommand{\url}[1]{#1}
\csname url@samestyle\endcsname
\providecommand{\newblock}{\relax}
\providecommand{\bibinfo}[2]{#2}
\providecommand{\BIBentrySTDinterwordspacing}{\spaceskip=0pt\relax}
\providecommand{\BIBentryALTinterwordstretchfactor}{4}
\providecommand{\BIBentryALTinterwordspacing}{\spaceskip=\fontdimen2\font plus
\BIBentryALTinterwordstretchfactor\fontdimen3\font minus \fontdimen4\font\relax}
\providecommand{\BIBforeignlanguage}[2]{{%
\expandafter\ifx\csname l@#1\endcsname\relax
\typeout{** WARNING: IEEEtran.bst: No hyphenation pattern has been}%
\typeout{** loaded for the language `#1'. Using the pattern for}%
\typeout{** the default language instead.}%
\else
\language=\csname l@#1\endcsname
\fi
#2}}
\providecommand{\BIBdecl}{\relax}
\BIBdecl

\bibitem{zfbf}
G.~Caire and S.~Shamai, ``On the achievable throughput of a multiantenna gaussian broadcast channel,'' \emph{IEEE Transactions on Information Theory}, vol.~49, no.~7, pp. 1691--1706, 2003.

\bibitem{wmmse}
Q.~Shi, M.~Razaviyayn, Z.~Q. Luo, and C.~He, ``{A}n iteratively weighted {MMSE} approach to distributed sum-utility maximization for a {MIMO} interfering broadcast channel,'' \emph{IEEE Trans. Signal Process.}, vol.~59, pp. 4331--4340, 2011.

\bibitem{ccfp}
K.~Shen and W.~Yu, ``Fractional programming for communication systems—part i: Power control and beamforming,'' \emph{IEEE Transactions on Signal Processing}, vol.~66, no.~10, pp. 2616--2630, 2018.

\bibitem{ergodic_stoch:kalogerias2020model}
D.~S. Kalogerias, M.~Eisen, G.~J. Pappas, and A.~Ribeiro, ``Model-free learning of optimal ergodic policies in wireless systems,'' \emph{IEEE Transactions on Signal Processing}, vol.~68, pp. 6272--6286, 2020.

\bibitem{lu2019robust:imperfectcsi}
A.-A. Lu, X.~Gao, W.~Zhong, C.~Xiao, and X.~Meng, ``Robust transmission for massive mimo downlink with imperfect csi,'' \emph{IEEE Transactions on Communications}, vol.~67, no.~8, pp. 5362--5376, 2019.

\bibitem{yoo2006optimality:opportunistic}
T.~Yoo and A.~Goldsmith, ``On the optimality of multiantenna broadcast scheduling using zero-forcing beamforming,'' \emph{IEEE Journal on selected areas in communications}, vol.~24, no.~3, pp. 528--541, 2006.

\bibitem{ribeiro2012optimal:opportunistic}
A.~Ribeiro, ``Optimal resource allocation in wireless communication and networking,'' \emph{EURASIP Journal on Wireless Communications and Networking}, vol. 2012, no.~1, p. 272, 2012.

\bibitem{dro:li2014}
Q.~Li, A.~M.-C. So, and W.-K. Ma, ``Distributionally robust chance-constrained transmit beamforming for multiuser miso downlink,'' in \emph{2014 IEEE International Conference on Acoustics, Speech and Signal Processing (ICASSP)}, 2014, pp. 3479--3483.

\bibitem{dro:li2017distributionally}
B.~Li, Y.~Rong, J.~Sun, and K.~L. Teo, ``A distributionally robust minimum variance beamformer design,'' \emph{IEEE Signal Processing Letters}, vol.~25, no.~1, pp. 105--109, 2017.

\bibitem{bazzi2023:outageprob}
A.~Bazzi and M.~Chafii, ``On outage-based beamforming design for dual-functional radar-communication 6g systems,'' \emph{IEEE Transactions on Wireless Communications}, vol.~22, no.~8, pp. 5598--5612, 2023.

\bibitem{alsenwi2023cvar:mincvarrate}
M.~Alsenwi, E.~Lagunas, H.~Al-Hraishawi, and S.~Chatzinotas, ``Cvar-based robust beamforming framework for massive mimo leo satellite communications,'' in \emph{GLOBECOM 2023-2023 IEEE Global Communications Conference}.\hskip 1em plus 0.5em minus 0.4em\relax IEEE, 2023, pp. 3953--3958.

\bibitem{li2024hpe:minqoscstr}
Y.~Li and Y.-F. Liu, ``Hpe transformer: Learning to optimize multi-group multicast beamforming under nonconvex qos constraints,'' \emph{IEEE Transactions on Communications}, vol.~72, no.~9, pp. 5581--5594, 2024.

\bibitem{outagecap:lifang2001}
L.~Li and A.~Goldsmith, ``Capacity and optimal resource allocation for fading broadcast channels .ii. outage capacity,'' \emph{IEEE Transactions on Information Theory}, vol.~47, no.~3, pp. 1103--1127, 2001.

\bibitem{risk:bennis2018}
M.~Bennis, M.~Debbah, and H.~V. Poor, ``Ultrareliable and low-latency wireless communication: Tail, risk, and scale,'' \emph{Proceedings of the IEEE}, vol. 106, no.~10, pp. 1834--1853, 2018.

\bibitem{risk:li2021}
Y.~Li, D.~Guo, Y.~Zhao, X.~Cao, and H.~Chen, ``Efficient risk-averse request allocation for multi-access edge computing,'' \emph{IEEE Communications Letters}, vol.~25, no.~2, pp. 533--537, 2021.

\bibitem{risk:vu2018}
T.~K. Vu, M.~Bennis, M.~Debbah, M.~Latva-aho, and C.~S. Hong, ``Ultra-reliable communication in 5g mmwave networks: A risk-sensitive approach,'' \emph{IEEE Communications Letters}, vol.~22, no.~4, pp. 708--711, 2018.

\bibitem{dro:yaylali2023}
G.~Yaylali and D.~Kalogerias, ``Robust and reliable stochastic resource allocation via tail waterfilling,'' in \emph{2023 IEEE 24th International Workshop on Signal Processing Advances in Wireless Communications (SPAWC)}, 2023, pp. 256--260.

\bibitem{dro:yaylali2024}
------, ``Distributionally robust power policies for wireless systems under power fluctuation risk,'' in \emph{2024 58th Asilomar Conference on Signals, Systems, and Computers}, 2024, pp. 1005--1012.

\bibitem{cvar:shapiro}
A.~Shapiro, D.~Dentcheva, and A.~Ruszczynski, \emph{{L}ectures on {S}tochastic {P}rogramming: {M}odeling and {T}heory}, 3rd~ed.\hskip 1em plus 0.5em minus 0.4em\relax SIAM, 2021.

\bibitem{cvar:rockafellar}
R.~T. Rockafellar, S.~Uryasev \emph{et~al.}, ``Optimization of conditional value-at-risk,'' \emph{Journal of risk}, vol.~2, pp. 21--42, 2000.

\bibitem{shlezinger2023model:MBDL}
N.~Shlezinger, J.~Whang, Y.~C. Eldar, and A.~G. Dimakis, ``Model-based deep learning,'' \emph{Proceedings of the IEEE}, vol. 111, no.~5, pp. 465--499, 2023.

\bibitem{GNN1}
F.~Scarselli, M.~Gori, A.~C. Tsoi, M.~Hagenbuchner, and G.~Monfardini, ``The graph neural network model,'' \emph{IEEE Transactions on Neural Networks}, vol.~20, no.~1, pp. 61--80, 2009.

\bibitem{zhao2024understanding:cnngnncomp}
B.~Zhao, J.~Guo, and C.~Yang, ``Understanding the performance of learning precoding policies with graph and convolutional neural networks,'' \emph{IEEE Transactions on Communications}, vol.~72, no.~9, pp. 5657--5673, 2024.

\bibitem{yang2024knowledge:uwgnn}
H.~Yang, N.~Cheng, R.~Sun, W.~Quan, R.~Chai, K.~Aldubaikhy, A.~Alqasir, and X.~Shen, ``Knowledge-driven resource allocation for wireless networks: A wmmse unrolled graph neural network approach,'' \emph{IEEE Internet of Things Journal}, 2024.

\bibitem{DavisDrusvyatskiyKakadeLee2020_TameSubgradient}
D.~Davis, D.~Drusvyatskiy, S.~M. Kakade, and J.~D. Lee, ``{S}tochastic subgradient method converges on tame functions,'' \emph{Foundations of Computational Mathematics}, vol.~20, no.~1, pp. 119--154, 2020.

\bibitem{BolteLeMoulinesPauwels2024_InexactSubgradient}
J.~Bolte, T.~Le, E.~Moulines, and E.~Pauwels, ``{I}nexact subgradient methods for semialgebraic functions,'' \emph{Mathematical Programming}, 2024.

\bibitem{sharperatio}
W.~F. Sharpe, ``The sharpe ratio,'' \emph{Streetwise--The Best of the Journal of Portfolio Management}, vol.~3, no.~3, pp. 169--85, 1998.

\end{thebibliography}

\end{document}